\documentclass[12pt]{article}

\usepackage[english]{babel}
\usepackage{standalone}
\usepackage[utf8x]{inputenc}
\usepackage[T1]{fontenc}
\usepackage[a4paper,top=2.5cm,bottom=2cm,left=2.2cm,right=2.2cm]{geometry}
\usepackage{amsmath, amssymb, amsthm, amsbsy, amsfonts, natbib,setspace}
\usepackage{graphicx, comment,xcolor, float, wrapfig, verbatim, pspicture, epsfig, rotating, subfigure,  colortbl, threeparttable}
\usepackage{indentfirst} 
\usepackage[colorinlistoftodos]{todonotes}
\usepackage[colorlinks=true, allcolors=blue]{hyperref}
\allowdisplaybreaks
\newcommand{\bs}{\boldsymbol}
\newcommand{\E}{\mbox{E}}
\newcommand{\y}{\mathbf{y}}
\newcommand{\x}{\mathbf{x}}
\newcommand{\dd}{\mathbf{d}}

\title{\large{\textbf{Assessment of Bayesian Expected Power via Bayesian Bootstrap}}}
\author{Fang Liu$^{\ast}$\\
\normalsize{Applied and Computational Mathematics and Statistics}\\
 \normalsize{ University of Notre Dame, Notre Dame, IN 46556}\\
\normalsize{$^\ast$\textit{email}: fang.liu.131@nd.edu}}
\date{\today}

\begin{document}
\maketitle
The Bayesian expected power (BEP) has become  increasingly popular in sample size determination and assessment of the probability of success (POS) for a future trial. The BEP  takes into consideration the uncertainty around the parameters assumed by a power analysis and is thus more robust compared to the traditional power that assumes a single set of parameters. Current methods for assessing BEP are often based in a parametric framework by imposing a model on the pilot data to derive and sample from  the posterior distributions of the parameters. Implementation of the model-based approaches can be  analytically challenging and computationally costly especially for multivariate data sets; it also runs the risk of generating misleading BEP if the model is mis-specified.  We propose an approach based on the Bayesian bootstrap technique (BBS)  to simulate future trials  in the presence of individual-level pilot data, based on which the empirical BEP can be calculated. The BBS approach  is model-free with no assumptions about the distribution of the prior data and circumvents the analytical and computational complexity associated with obtaining the posterior distribution of the parameters. Information from multiple pilot studies is also straightforward  to combine. We also propose the double bootstrap (BS2), a frequentist counterpart to the BBS, that shares similar properties and achieves the same goal as the BBS for BEP assessment. Simulation studies and case studies are presented  to demonstrate the implementation of the BBS and BS2 techniques and to compare the BEP results with model-based approaches.
\vspace{12pt}


\newpage

\section{Introduction}\label{sec:introduction}
The probability of success (POS) of a clinical trial is affected by many factors such as patient  recruitment, ethical considerations,  local regulations, resources, study designs and execution, among  others. The part where statisticians get involved the most  is the study design during the trial planning stage, including the sample size determination and power calculation.  Power is defined as the probability of rejecting the null hypothesis  $H_0$ if the alternative hypothesis $H_1$  is true in a future trial. The probabilistic nature of power makes a natural choice as a metric for measuring how the statistical design aspect of the study affects the POS of the trial. On the other hand, the classical  power is a conditional probability given a specific effect size $\Delta=\mu/\sigma$, the value of which is an unknown quantity and is often the parameter of primary interest  to estimate in the planned trial. As such, the power  is sensitive to the assumed effect size and not a robust measurement of POS.

A more robust measurement for POS compared to the classical power is the  Bayesian expected power (BEP) or the weighted average power (WAP), which is the expected power over the distribution of the effect size $\Delta$ given existing data. In other words, the BEP is a ``marginal'' measure of the POS given what's known, integrating out the unknown  $\Delta$. The BEP has been ``re-invented'' several times under different names in various contexts.   \citet{brown} suggested using Bayesian methods to obtain a posterior distribution representing the state of knowledge of the parameters of interest, and to predict the outcome of a specified comparative trial. The approach  was implemented and demonstrated in several examples in \citet{spiegel}.  \citet{ohagan2001} and \citet{ohagan2005} used the term ``assurance'' and \citet{chuang} used ``average success probability'' to describe the BEP.   In all thedr cases, only  the uncertainty around $\mu$ concerned by the hypothesis in the future trial is considered and the uncertainty around the variance $\sigma$, which is an equally important parameter in power calculation, is not taken into consideration. \citet{liu} extended the BEP by accounting for the uncertainty around both $\mu$ and $\sigma$ (or the effect size $\Delta$), and also added two more versions of the BEP to by removing the ``type I error''  component  from the regular BEP metric. Some recent reviews, discussion, and applications of the BEP in early and late phased clinical trials, and in meta-analysis are given in \citet{chuang2012, carroll, ibrahim, han,bliss2016}. The BEP has been  routinely calculated alongside  the traditional power in some pharmaceutical companies; and it is also implemented in several  sample size and power calculation software \citep{r,cytel}. 

The current approaches to assessing the BEP often start with constructing the joint posterior distributions of $\mu$ and $\sigma$ given a prior and the likelihood of the parameters given the pilot/exisiting data. In many cases, the posterior distributions of $\mu$ and $\sigma$ may not have closed-form expressions, or even they do, iterative approaches, such as the MCMC algorithms  and other sampling techniques, might still be necessary to draw the parameters from the posterior distributions. In the case when there are co-primary hypotheses,  $\boldsymbol{\mu}$ is multi-dimensional, analytically and computational it becomes even harder. In summary, the model-based approaches can be  analytically challenging and computationally costly especially for multivariate data.   On top of all these, there is always the risk of misspecifying the likelihood of the parameters with the pilot data, leading to misleading BEP values subsequently. 

We propose  an approach based on the Bayesian bootstrap technique \citep{rubin}  , referred to as the BBS approach, to assess the BEP when  individual-level pilot data  $\y$  are available. The Bayesian bootstrap is a Bayesian version of the bootstrap technique \citep{efron}. \citet{rubin}   proved that the Bayesian bootstrap is operationally and inferentially similar to the frequentist bootstrap.  In the BBS approach, we repeatedly simulate the future trial data given $\y$  via the Bayesian bootstrap technique without imposing any distributional assumptions on $\y$, and test $H_0$ according to the planned analysis in each set of the simulated trials; and  the overall rejection rate of $H_0$ over the repetitions leads to a Monte Carlo (MC) estimate of the BEP. The uncertainty around the underlying true distribution of the  pilot data is accounted for by placing a  Dirichlet prior on the probability of each individual. The BBS approach is straightforward to implement with minimal analytical work except for the planned analysis on the simulated future data. Computationally, only a few lines of codes are needed to implement the sampling and trial simulation steps, and the whole BBS procedure can be easily parallelized for fast computation.  When there are multiple  relevant pilot data sets, it is straightforward to combine the pilot data together via the BBS approach.

Directly applying the regular bootstrap to the pilot data to simulate future data is inappropriate for the purpose of BEP assessment since it will not  propagate the uncertainty around the distribution of the pilot data. As a matter of fact, the overall rejection rate based on the future data simulated this way is a MC estimate of the classical power assuming the estimated effect size from the pilot data is the true effect size. This motivates us to  come up with the ``double bootstrap'' technique (BS2), a frequentist counterpart to the BBS, that will propagate the uncertainty around the distribution of the pilot data and achieves the same goal as the BBS for BEP assessment. Procedurally, the BS2 first bootstraps the pilot data, and then samples the bootstrapped pilot data to generate repetitions for the future trials based on which a MC estimate of the BEP is obtained. Similar to the BBS, only a few lines of codes are needed to implement the BS2, and it is easy to parallelize computationally.
 
The rest of the paper is organized as follows. Section \ref{sec:method} introduces the BBS and BS2 approaches for assessing the BEP. Section \ref{sec:simulation} compares the BBS and BS2 approaches with the parametric approaches in the BEP assessment in two simulation studies. Section \ref{sec:example} implements the BBS and BS2 approaches to assess BEP for an equivalence study and for a HIV survival study. Section \ref{sec:discussion} concludes the discussion with some final remarks. The R codes for the numerical examples in Section \ref{sec:example} are provided in the online supplementary materials to this paper and are also available for download at \url{TBD}.

\section{Methods}\label{sec:method}

\subsection{assessment of BEP}
By definition, the BEP is the probability of rejecting $H_0$ in a future trial given existing data $\y$. Denote the classical power by $\beta(\bs{\theta})=\Pr$(rejecting $H_0|\bs{\theta},\tilde{n}$), where $\bs{\theta}$ refers to the parameters involved in the power calculation and  $\tilde{n}$ is the given sample size of the future trial, then
\begin{equation}
\mbox{BEP}=\E[\beta(\bs{\theta})|y]=\textstyle\int\beta({\bs{\theta}})p(\bs{\theta}|\y)d\bs{\theta}.\label{eqn:old}
\end{equation}
To calculate the BEP, we may first draw $\bs{\theta}$ from its posterior distribution $p(\bs{\theta}|\y)$, and then plug in the drawn $\bs{\theta}$  in the power function $\beta(\bs{\theta})$ to obtain a classical power value given the drawn $\bs{\theta}$. Repeating the two steps many times, say $m$, will lead to the posterior distribution of the power $\beta$ given $\y$, the average $\sum_{j=1}^m \beta(\bs{\theta^{(j)}})$ of the posterior samples is an MC estimate for BEP for the future trial for a given $\tilde{n}$. Besides the BEP, other statistics from the posterior distribution of the power can also be reported, such as the mode and the percentiles.

If a closed-formed function $\beta(\bs{\theta})$  is not available, the MC method can be applied to numerically approximate the power by simulating the future trial data $\tilde{\y}$ on which $H_0$ will be tested. Denote the event of rejecting $H_0$ in the future trial by $I(R(\tilde{\y}))$, where $R$ is the rejection rule, a function of $\tilde{\y}$, and $I(R(\tilde{\y}))=1$ if $H_0$ is rejected in $\tilde{\y}$  and 0 otherwise, then
\begin{eqnarray}
\mbox{BEP}&=& \E\left[I(R(\tilde{\y}))|\y\right]=\E\left[\E[I(R(\tilde{\y}))|\bs{\theta},\y]|\y\right]\notag\\
&=&\textstyle\int\!\int\! I(R(\tilde{\y}))p(\tilde{\y}|\y,\bs{\theta})p(\bs{\theta}|\y) d\tilde{\y}d\bs{\theta}\notag\\
&=&\textstyle\int\!\left(\int\! I(R(\tilde{\y}))p(\tilde{\y}|\bs{\theta})d\tilde{\y}\right) p(\bs{\theta}|\y)d\bs{\theta}.\label{eqn:new}
\end{eqnarray}
Eqn (\ref{eqn:new})  suggests that we can first draw $\bs{\theta}$ from the posterior distribution $p(\bs{\theta}|\y)$ given pilot data $\y$, and simulate the future data $\tilde{\y}$ of size $\tilde{n}$ given the drawn $\bs{\theta}$, then perform the planned statistical analysis and hypothesis testing on the simulated data $\tilde{\y}$, and record 1 if rejected and 0 if not.  Repeating the process many times, say $m$, and the rejection rate over the $m$ repetations is an MC estimate of the BEP. We refer to this MC approach for assessing the BEP as the  future trial simulation (FTS) approach. 

Regardless of whether the model-based approach is employed when closed-form  $\beta(\bs{\theta})$ is available or the FTS approach is applied when it is not, the current approaches for the BEP assessment are mainly model-based by imposing distributional or model assumptions on the pilot data $\y$ to obtain the posterior distribution $p(\bs{\theta}|\y)$, followed by the posterior sampling step. If $\tilde{\y}$ and $\bs{\theta}$ are multidimensional, the sampling  of $\bs{\theta}$  and simulation of $\tilde{\y}$ can be computationally costly, not to be mention the risk of mis-specification of the parametric model, leading to misleading BEP subsequently.  Alternative techniques that avoid making strong parametric assumptions about data $\y$ and at the same time are computationally less complicated are desired.

We propose a FTS technique, referred to as the BBS approach, motivated by the Bayesian bootstrap to evaluate the BEP numerically. The BBS assesses the BEP without imposing a model on the pilot data or sampling from complicated posterior distributions via MCMC approaches. 
Instead, the BBS approach obtains the posterior distributions of the population distribution underlying $\y$ given $\y$, from which the future data will be simulated. We will also present the frequentist version of the BBS approach -- the ``double bootstrap'' technique (BS2), and discuss why the regular bootstrap will only lead to a MC estimation of the classical power instead of yielding the BEP.

\subsection{construction of a posterior distribution of the population distribution via Bayesian bootstrap}\label{sec:bayesian.bootstrap}
Define a finite population of size $N$ with $K$ ($K<\infty$) distinct values over $p$ attributes.  Let $\pi_k$ denote the probability that value $\dd_k$ occurs for $k=1,\ldots,K$. The Bayesian bootstrap can be used to obtain a posterior distribution of the population distribution $f$ of the $K$ distinct values given a data set $\y$ of size $n$. Denote by $\bs{\pi}=(\pi_1,\ldots,\pi_K)$ the probabilities associated with the $K$ distinct values ($\sum_k\pi_k=1$). Since $\bs{\pi}$ fully characterizes  $f$, obtaining the posterior distribution on $f$ is equivalent to obtaining the posterior distribution of $\bs{\pi}$.  A convenient choice on the prior of $\bs{\pi}$ is the Dirichlet distribution $\mbox{D}(\bs{\pi}| \alpha_1,\ldots,\alpha_K)\propto\prod_{k=1}^K\pi_k^{\alpha_k-1}$, with hyper-parameters $\bs{\alpha}=(\alpha_1,\ldots,\alpha_K)>0$. Since Dirichlet priors are conjugate priors for the multinomial likelihood, the posterior distribution of $\bs{\pi}$ given data $\y$ (of size $n$) also follows a Dirichlet distribution $\mbox{D}(\pi_1,\ldots,\pi_{K}|\mathbf{n})\propto\prod_{k=1}^{K}\pi_k^{n_k+\alpha_k-1}$, where  $\mathbf{n}=(n_1,\ldots, n_K)$, $n_k$ is the number of observations in the sample that take the value $\dd_k$, and $\sum_k n_k=n$.

It is highly likely that not every distinct  $\dd_k$ from the population will occur in the sample; and $n_k$'s associated with these ``non-appearing'' cases will be $0$.  Denote by $L$ the number of distinct cases in the sample ($L\le K$), the posterior distribution of the subset of $\bs{\pi}$ that is associated with the sample cases is $D(\pi_1,\ldots,\pi_{L}|\mathbf{n})\propto\prod_{k=1}^{L}\pi_k^{n_k+\alpha_k-1}$. In practical application, we can always set $L=n$ due to two reasons: it does not  affect the posterior distribution of $\bs{\pi}$ associated with the distinct values  given the aggregation property of the Dirichlet distribution; and every individual in the sample can be distinct from each other when there are continuous variables among the $p$ attributes or $p$ is large.  With $L=n$, $n_k=1$ for $k=1,\ldots, n$, and  The posterior distribution of $\bs{\pi}$ can be simplified to
\begin{equation}\label{eqn:post1}
p(\pi_1,\ldots,\pi_n|\y)=\frac{\prod_{i=1}^n \pi_k^{(\alpha_k+1)-1}}{B(\alpha_1+1,\ldots,\alpha_n+1)}=\mbox{D}(\alpha_1+1,\ldots,\alpha_n+1).
\end{equation}
In terms of the choice for the hyper-parameters $\alpha_k$, proper priors should have $\alpha_k>0\;\forall\;k=1,\ldots,n$.  If there are minimal information about $\bs{\pi}$ prior to the pilot data $\y$, setting $\alpha_k$ at small positive numbers  (eg, 0.1, 0.5, 1, etc) leads to weakly informative priors for $\bs{\pi}$. An improper non-informative but convenient choice is $\alpha_k=0\;\forall\;k=1,\ldots,n$, which  still generates a proper posterior distribution for $\bs{\pi}$,
\begin{equation}\label{eqn:post2}
p(\pi_1,\ldots,\pi_n|\y)=\left(\Gamma(n)\right)^{-1}=\mbox{D}(1,\ldots,1).
\end{equation}

\subsection{BEP assessment with Bayesian bootstrap}\label{sec:bbs}
The Bayesian bootstrap is employed in the BEP assessment to  simulate data $\tilde{\y}$ for the future trial,  the posterior predictive distribution of which, given pilot data $\y$, is
\begin{equation}\label{eqn:ppd}
p(\tilde{\y}|\y)=\textstyle\int p(\tilde{\y}|\y,\bs{\pi})p(\bs{\pi}|\y)d\bs{\pi}
=\textstyle\int p(\tilde{\y}|\bs{\pi})p(\bs{\pi}|\y)d\bs{\pi}=\textstyle\int\! M(\tilde{n},\bs{\pi})p(\bs{\pi}|\y)d\bs{\pi},
\end{equation}
where $p(\bs{\pi}|\y)$ is given in Eqn (\ref{eqn:post1}) (Eqn (\ref{eqn:post2}) if $\alpha_k=0$).  Plugging in $p(\tilde{\y}|\y)$ from Eqn (\ref{eqn:ppd}) in Eqn (\ref{eqn:new}), we have

\begin{equation}\label{eqn:new2}
\mbox{BEP}
=\textstyle\int\!\!\int I(R(\tilde{\y})) M(\tilde{n},\bs{\pi})p(\bs{\pi}|\y)d\bs{\pi}d\tilde{\y}
\end{equation}
Eqn (\ref{eqn:new2}) suggests that BEP can be calculated via the steps given in  Table \ref{tab:bbs1}, which produces a single metric of BEP rather than the posterior distribution of power. If it is desired to have the posterior distribution of power, an inner loop will need to be built into the BBS procedure, as given in Table \ref{tab:bbs2}.  When the number of iterations of the inner loop $t=1$, Table \ref{tab:bbs2} reduces to Table \ref{tab:bbs1}.
\begin{table}[!htb]
\begin{tabular}{l}
\hline
\textbf{DO} $j=1,\ldots, m$\\
1) draw $\bs{\pi}^{(j)}$ from the Dirichlet distribution in Eqn (\ref{eqn:post1});\\
2) draw $\tilde{\y}^{(j)}$ from  $\mbox{M}(\tilde{n}, \bs{\pi}^{(j)})$;\\
3) test $H_0$ in simulated trial $\tilde{\y}^{(j)}$ and record $I(R(\tilde{\y}^{(j)}))$ (1 if $H_0$ is rejected; 0 otherwise); \\
\textbf{END DO}\\
\textbf{OUTPUT}: calculate BEP $=m^{-1}\sum_{j=1}^m I(R(\tilde{\y}^{(j)}))$. \\
\hline
\end{tabular}
\caption{BEP assessment  via the BBS}\label{tab:bbs1}
\end{table}

 \begin{table}[!htb]
\begin{tabular}{l}
\hline
\textbf{DO} $j=1,\ldots, m$\\
1) draw $\bs{\pi}^{(j)}$ from the Dirichlet distribution in Eqn (\ref{eqn:post1});\\
2) given the drawn  $\bs{\pi}^{(j)}$\\
\hspace{6pt} \textbf{DO} $l=1,\ldots, t$\\
\hspace{6pt}  2.1) draw $\tilde{\y}^{(l,j)}$  from  $\mbox{M}(\tilde{n}, \bs{\pi}^{(j)})$;\\
\hspace{6pt}  2.2) test $H_0$ in simulated trial $\tilde{\y}^{(l,j)}\!$ and record $I(R(\tilde{\y}^{(l,j)}))\!$ (1 if $H_0$ is rejected; 0 o.w.); \\
\hspace{6pt}  \textbf{END DO}\\
3) calculate power $\beta(\bs{\pi}^{(j)})=t^{-1}\sum_{l=1}^t I(R(\tilde{\y}^{(l,j)}))$.\\
\textbf{END DO}\\
\textbf{OUTPUT}:  $m$ samples from the posterior distribution of power: $(\beta(\bs{\pi}^{(1)}),\ldots,\beta(\bs{\pi}^{(m)}))$\\
\hspace{58pt}   and  BEP $=m^{-1}\sum_{j=1}^m(\beta(\bs{\pi}^{(j)})$.\\
\hline
\end{tabular}
\caption{Generation of the posterior distribution of power via the BBS}\label{tab:bbs2}
\end{table}

\subsection{the double boostrap}\label{sec:bs2}
The BBS procedures in Tables \ref{tab:bbs1} and \ref{tab:bbs2} for assessing the BEP are based in the framework of the Bayesian bootstrap. The frequentist counterpart to the BBS approach  is the ``double bootstrap'' (BS2)  given in Table \ref{tab:bbs3}. The outer-loop bootstrap captures the uncertainty around the population distribution (or the population parameters), corresponding to the sampling of $\bs{\pi}$ from its posterior distribution in the BBS procedure. The inner-loop bootstrap propagates the sampling variability and error for the future trial, serving the same purposes of drawing a sample data set given $\bs{\pi}$ in the BBS procedure. The BS2 technique outputs samples from a  distribution of power given pilot data $y$, the average of which gives the WAP, taking into account the uncertainty of the sample data.   When $m\rightarrow\infty$, the BS2 technique is equivalent to the BBS approach when the hyper-parameters $\alpha_k=0\;\forall\; k=1,\ldots,n$.  Strictly speaking,  the  BS2 is not Bayesian conceptually, and the distribution of the power generated from the procedure is thus not a posterior distribution though it is a conditional distribution given the pilot data $\y$, and the mean of which is asymptotically equivalent to the BEP. Setting $t=1$ for the inner loop in Table \ref{tab:bbs3} yields a single WAP estimate without  a distribution of power (similar to Table  \ref{tab:bbs2} being reduced to Table \ref{tab:bbs1} in the BBS).
\begin{table}[!htb]
\begin{tabular}{l}
\hline
\textbf{DO} $j=1,\ldots, m$\\
1) bootstrap a sample $S^{(j)}$ of size $n$ from $\y$ with replacement;\\
\hspace{6pt} \textbf{DO} $l=1,\ldots, t$\\
\hspace{6pt}  2.1) bootstrap $\tilde{\y}^{(l,j)}$ of size $\tilde{n}$ from $S^{(j)}$ with replacement;\\
\hspace{6pt}  2.2) test $H_0$ in simulated trial $\tilde{\y}^{(l,j)}\!$ and record $I(R(\tilde{\y}^{(l,j)}))\!$ (1 if $H_0$ is rejected; 0 o.w.); \\
\hspace{6pt}  \textbf{END DO}\\
3) calculate power $\beta(\bs{\pi}^{(j)})=t^{-1}\sum_{l=1}^t I(R(\tilde{\y}^{(l,j)}))$.\\
\textbf{END DO}\\
\textbf{OUTPUT}: $m$ samples from the conditional distribution of power given $\y$:  \\
\hspace{58pt}   $(\beta(\bs{\pi}^{(1)}),\ldots,\beta(\bs{\pi}^{(m)}))$,  and  EP $=m^{-1}\sum_{j=1}^m(\beta(\bs{\pi}^{(j)})$.\\
\hline
\end{tabular}
\caption{Generation of the conditional distribution of power given pilot data via the BS2}\label{tab:bbs3}
\end{table}

If we directly simulate the future data  $\tilde{\y}$ via the regular bootstrap, we end up having a procedure (Table \ref{tab:power}) that leads to a MC estimate of the classical power assuming what's observed in the pilot data is the truth.
\begin{table}[!htb]\begin{center}
\begin{tabular}{l}
\hline
\textbf{DO} $j=1,\ldots, m$\\
1) bootstrap  $\tilde{\y}^{(j)}$ of size $\tilde{n}$ from $\y$ with replacement;\\
2) test $H_0$ in $\tilde{\y}^{(j)}$  and record $I(R(\tilde{\y}^{(j)}))\!$ (1 if $H_0$ is rejected; 0 o.w.); \\
\textbf{END DO}\\
\textbf{OUTPUT}: power $=m^{-1}\sum_{j=1}^m I(R(\tilde{\y}^{(j)}))$.\\
\hline
\end{tabular}
\end{center}\caption{Monte Carlo power via FTS given pilot data}\label{tab:power}
\end{table}\
It is obvious that this procedure does not take account the uncertainty around the unknown population distribution underlying $\y$, thus implicitly assumes the observed pilot data is the whole population and the estimated effect size from $\y$ is the true effect size of the population.

\section{Simulation Studies}\label{sec:simulation}
We run two simulation studies to implement the BBS and BS2 procedures in Tables \ref{tab:bbs2} and \ref{tab:bbs3} to assess the BEP, and compared the results to the model-based  BEP with an assumed model on the pilot data. We also computed the classical power  we compared the MC power via the bootstrap procedure  in Table \ref{tab:power}  to the model-based power assuming the observed results from the pilot study is the truth. The type I error rates in the hypothesis testing in future trials were 5\% in both simulation studies.

\begin{figure}[!htb]
\includegraphics[height=0.52\textheight,width=1.0\textwidth]{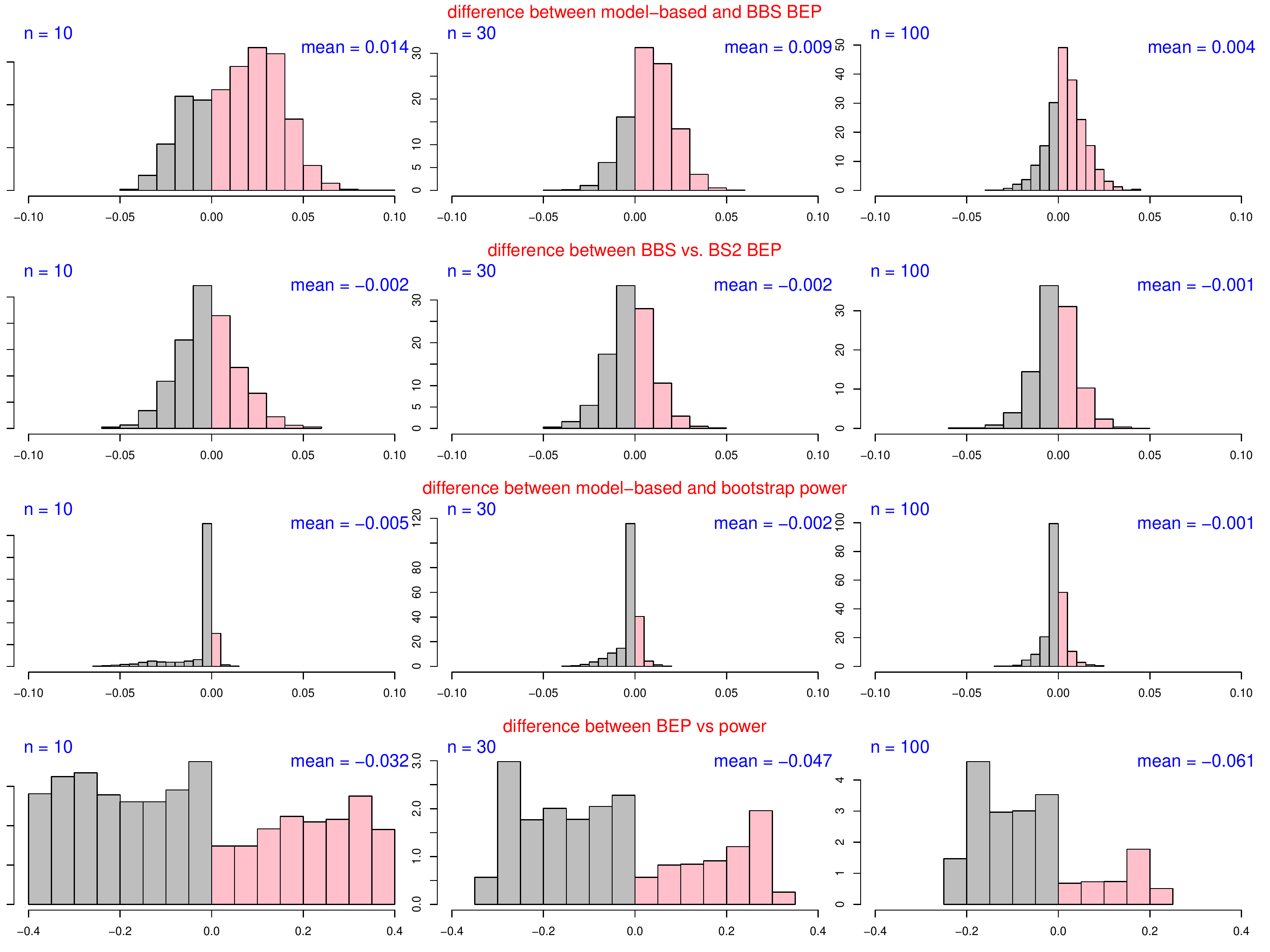}
\caption{Histogram of the differences between two metrics (model-based vs BBS BEP; BBS vs BS2 BEP; model-based vs bootstrap power; and power vs BEP over 2500 repetitions in Simulation 1 (gray bars: $A<B$ in the $A$ vs $B$ comparison, pink bars: power $A>B$)}\label{fig:sim1diff}
\end{figure}
In the first simulation study, 2500 repetitions were run. In each repetition, a pilot of study of size $n$  was simulated from N$(\mu=0.15,\sigma=1)$. We examined 3 cases of $n$: $n=$10, 30 and 100, respectively. The future trial under planning had $\tilde{n}=500$ and the hypotheses were $H_0\!:\!\mu=0$ vs $H_1\!:\!\mu\ne 0$ . The power assuming the effect size from the pilot study was true and the BEP were calculated analytically and via the bootstrap procedures.   Figure \ref{fig:sim1diff} depicts the distributions for the  differences  over the 2500 repetitions between the various metrics, and the results are summarized as follows. First, the BEP calculated analytically and via the BBS and BS2  was similar; the small discrepancy between the two decreased with $n$ (the first and second rows). Second, the power calculated via the bootstrap procedure was similar to the model-based power; the small discrepancy between the two decreased with the pilot study size $n$ (the first row). Third, the BEP could be larger or smaller than power, depending on the pilot data. The mean difference between the two increased with $n$ and the dispersion of the difference decreased  (the fourth row).  From the histograms and the empirical CDFs (cumulative distribution functions) of the power and BEP presented in the online supplementary materials, we observe that the CDFs of the power and the BEP intersected when power and BEP around $0.5$ regardless of $n$.  As $n$ increased, the distribution of the power became less ''bi-polar'' while that of the BEP became less uniform, and the two distributions became more similar. 

\begin{figure}[!htb]
\includegraphics[height=0.75\textheight,width=0.9\textwidth]{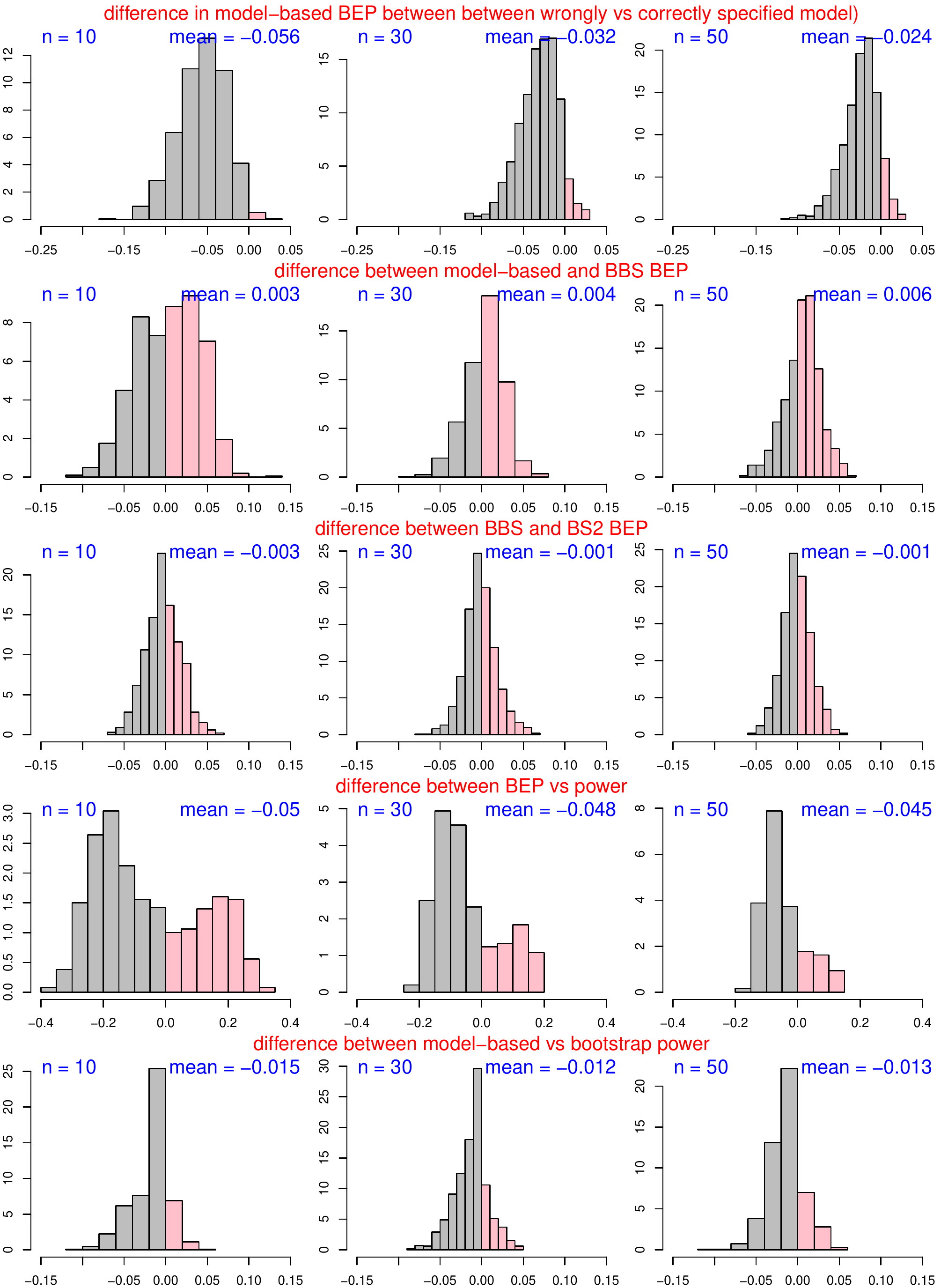}
\caption{Histogram of the differences between two metrics (model-based BEP based wrong vs correct models on the pilot data;  model-based vs BBS BEP; BBS vs BS2 BEP; model-based vs bootstrap power; and power vs BEP over 1000 repetitions in Simulation 2 (gray bars: $A<B$ in the $A$ vs $B$ comparison, pink bars: power $A>B$)}\label{fig:sim2diff}
\end{figure}
In the second simulation study, 1000 repetitions were run. In each repetition, a pilot of study of size $n$  was simulated from a bivariate lognormal distribution LN$(\bs{\mu}=(3,5)',\Sigma=(^1_{0.65} \;^{0.65}_1))$. We examined 3 cases of $n$: 10, 30, and 50. The future study had $\tilde{n}=80$ and the hypotheses were $H_0\!:\!(\mu_1\!\le\!2.7)\cup(\mu_2\!\le\!4.5)$ vs.  $\!\!H_1\!:\!(\mu_1\!>\!2.7)\cap(\mu_2\!>\!4.5)$.
Power assuming the effect size from the pilot study was true was calculated both analytically and via the bootstrap procedure, and the model-based BEP with a correctly specified model on the pilot data (a lognormal distribution) and with a wrongly  specified model (a normal distribution), and the BEP via the BBS and BS2 procedures  were calculated.  Figure \ref{fig:sim2diff} depicts the distributions of the  differences over the 1000 repetitions between the metrics.   The histograms of the metrics (2 powers, and 4 BEPs) over the 1000 simulations and  their empirical CDFs are available in the online supplementary materials. The results on the comparisons between the model-based (correctly specified) vs  bootstrap power, between the model-based (correctly specified), BBS, and BS2 BEP, and between the power and BEP are similar to those in Simulation 1. In terms of the model-based BEP based on the correctly specified model vs the wrongly specified model on the pilot data,  Figure \ref{fig:sim2diff} (the first row) suggests the BEP based on the wrong model was smaller than the correct BEP by 2.4\%  to 5.6\% on average, with larger $n$ blunting the effect of the model mis-specification on the BEP assessment.

\section{Cast Studies}\label{sec:example}
We implemented the BBS and the BS2 procedures in  two case studies. The future trial in the first example is a 2-period crossover study with an equivalence hypothesis and the pilot study is a 3-period crossover study. In the second example, the future trial is a two-arm study comparing the survival time of AIDS patients on two different treatments via the joint modelling of the survival data and the longitudinal data on CD4 counts; and the pilot study is of the same design but is smaller in size. The type I error rates in both examples are 5\%.

\subsection{example 1: crossover study with equivalence hypothesis}
A 2-period crossover study (sample size $\tilde{n}=200$) is under planning with the goal of developing a new formulation for a fixed-dose combination (FDC) drug for treating dislipidemia that has similar pharmacokinetic (PK) profile as a reference formulation R. The $H_1$ states that the geometric mean ratios (GMRs) between the  FDC and  R are within the interval of (0.80, 1.25) on 4 PK endpoints: 2 measures on the area under the PK curve (AUC)  and 2 measures of the maximum concentration (Cmax) on two chemical entities. There exists  a small pilot 3-period crossover study ($n=36$) with two candidate  FDC formulations FDC1 and FDC2, and the reference formulation R.  The team aims to use the test formulation  with the higher POS in the planned trial based on the information collected in the pilot study.

 When modelling  a single normally distributed endpoint $\y_i=(y_{i1},\ldots,y_{ip})^T$ from a crossover study with $p$ repeated measures, the linear mixed-effects (lme)  model $\y_i=\x_i\bs{\beta}+\mathbf{z}_i\bs{\gamma}_i+\epsilon_i$ is often used.  The fixed-effects term  $\bs{\beta}$  often includes an intercept, treatment effects,  and period effects. The random effects term $\bs{\gamma}_i\sim \mbox{N}(0,\mathbf{G})$ and the error term $\epsilon_i\sim \mbox{N}(0,R)$ together define the variance/covariance V$(\y_i)= \Sigma=\mathbf{z}_iG\mathbf{z}_i+R$; and $\Sigma$ can be as simple as ``compound  symmetry'' (CS) structure (constant marginal variance on the diagonal and constant covariance on all the off-diagonals) or as complex as ``unstructured'' (UN) (fully parameterized with $p(p+1)/2$ parameters). 

In the model-based assessment of BEP, we first obtained the posterior distribution of the treatment differences (on the log-scale) between T1 and R, and between T2 and R given the pilot data (AUC and Cmax in BE studies are often analyzed on the log scale). The hypothesis to be established in the future trial is a union of 4  BE hypotheses, one per primary endpoint. Therefore, the power is the joint probability that  the 95\% CIs for the 4 GMRs fell within the interval of (0.80, 1.25) simultaneously. To model the dependency structure among the 4 endpoints, we analyzed all 4 endpoints in one lme model with a fully parameterized covariance matrix $\Sigma_{12\times12}$. The likelihood function of $(\bs{\beta},\Sigma)$ was
\begin{equation}\label{eqn:l1}
L(\bs{\beta},\Sigma;\y,\mathbf{x})= \textstyle\prod_{i=1}^n\left(2\pi|\Sigma|\right)^{-1/2}\exp\{(\y_i-\x'_i\bs{\beta})'\Sigma^{-1}(\y_i-\x'_i\bs{\beta})\}.
\end{equation}
We imposed the prior $p(\bs{\beta},\Sigma)\!\propto\!|\Sigma|^{-(\nu_0+p+1)/2}\!$ on $\Sigma$, which is an inverse Wishart distribution  with the \emph{a priori}  degrees of freedom $\nu_0$, scale matrix $|\Sigma_0|\rightarrow0$, and $p=12$. We tried two different $\nu_0$ at $0$ and $p+1$, respectively, to examine the impact of different parametric assumptions on the model-based BEP. $\nu_0=0$ corresponds to the (improper) Jeffreys prior. When  $\nu_0=p+1$, the conditional posterior mean $\mbox{E}(\Sigma|\bs{\beta},\y)$ is the same  as the MLE for $\Sigma$ if $\bs{\beta}$ is known. The joint posterior distribution of $(\bs{\beta},\Sigma)$ was
\begin{equation}\label{eqn:p1}
p(\bs{\beta},\Sigma|\y,\mathbf{x})\propto |\Sigma|^{-(\nu_0+p+1+n)/2}\exp\{(\y_i-\mathbf{x}'_i\bs{\beta})'\Sigma^{-1}(\y_i-\mathbf{x}'_i\bs{\beta})\}.
\end{equation}
In the pilot study, 5 individuals  out of 36 had missing values (missing at random) from at least one  periods. To draw $(\bs{\beta},\Sigma)$ from their posterior distribution, we used the imputation-posterior (IP) algorithm and the Gibbs sampler by imputing $\y_{\mbox{\small{mis}}}$ given $\Sigma$ and $\bs{\beta}$ (the imputation step in Eqn (\ref{eqn:ymis})) and drawing $\Sigma$ and $\bs{\beta}$ respectively from their full conditional posterior distributions given $\y_{\mbox{\small{obs}}}$ and the imputed $\y_{\mbox{\small{mis}}}$ (the posterior step in Eqns (\ref{eqn:Sigma}) and (\ref{eqn:beta})). The technical details on the derivation of the equations are provided in the Appendix.
\begin{align}
& p(\y_{\mbox{\small{mis}}}|\bs{\beta},\Omega, \y_{\mbox{\small{obs}}})
=   N\left(\mathbf{x}_{\mbox{\small{mis}}}\bs{\beta},M\Omega M'\right),\label{eqn:ymis}\\
&p(\Sigma|\bs{\beta},\y_{\mbox{\small{mis}}}, \y_{\mbox{\small{obs}}})
=\mbox{Inv-Wishart}(\textstyle\sum_{i=1}^n \mathbf{e}_i\mathbf{e}'_i, \nu_0+n)\notag\\
=\;&\frac{|\Sigma|^{-(\nu_0+n+p+1)/2}|\sum_{i=1}^n\mathbf{e}_i\mathbf{e}'_i|^{n/2}
\mbox{exp}\left\{-\mbox{tr}\left(\sum_{i=1}^n\mathbf{e}_i\mathbf{e}'_i\Sigma^{-1}\right)/2\right\}}{2^{(n\times p/2)}\Gamma_p(n/2)},\label{eqn:Sigma}\\
&p(\bs{\beta}|\Omega,\y_{\mbox{\small{mis}}}, \y_{\mbox{\small{obs}}})
=    N\left((\mathbf{x}'\Omega^{-1}\mathbf{x})^{-1}\mathbf{x}' \Omega^{-1}\y, (\mathbf{x}'\Omega^{-1}\mathbf{x})^{-1}\right).\label{eqn:beta}
\end{align}
 $M_{np\times np}$ in Eqn (\ref{eqn:ymis}) was a diagonal indicator matrix of missingness on each observation with $m_{jj}=1$ if the $j\textsuperscript{th}$ response was missing and 0 otherwise for $j=1,\ldots, np$; $\sum_{i=1}^{n}\mathbf{e}_i\mathbf{e}'_i=\sum_{i=1}^{n}(\y_i-\mathbf{x}_i\bs{\beta})(\y_i-\mathbf{x}_i\bs{\beta})'$ in Eqn (\ref{eqn:Sigma}) was the scale matrix of the inverse Wishart distribution and $2\nu_0+n-p-1$ was the degree of freedom; $\y=(\y_{\mbox{\small{obs}}}, \y_{\mbox{\small{mis}}})$ and $\Omega$ was the block diagonal matrix with $n$ blocks of $\Sigma$ in Eqn (\ref{eqn:beta}).  Upon the convergence of the IP algorithm, after the burn-in and thinning periods, 1000 sets of posterior samples of $(\Sigma,\bs{\beta})$ were kept, based on which, 1000 power values were calculated analytically for the future trial via a multivariate $t$ distribution that incorporated the dependency structure among the 4 endpoints (refer to the online supplementary materials for the R function on the power calculation). The average of the 1000 power values yielded the model-based BEP.

The BEP assessment via the BBS and the BS2 techniques was much more straightforward than the model-based approach. Not only they involved minimal analytical work, coding of the procedures was also easy. After the steps in Table \ref{tab:bbs2} and  Table \ref{tab:bbs3} ($m=1000$ and $t=1$),  we tested the 4 equivalence hypotheses via a lme model by endpoint  in each of the simulated trials via the BBS and BS2 procedures. If the 95\% Cls of the log treatment differences  for all 4 endpoints  fell within $(\log(0.8),\log(1.25))$, the future trial was claimed a success.  The success rate out of the 1000 simulated trials  was the BEP.  

We also computed the classical power analytically and via the bootstrap approach (Table \ref{tab:power}). In the model-based approach, the log-scaled treatment differences in the 4 endpoints and the variance/covariance structure $\Sigma_{12\times12}$ were estimated from the pilot study by fitting a lme model. We modelled $\Sigma$ with two different structures to examine the sensitivity of the model-based power to the assumed model: a UN $\Sigma$ and a Kronecker product $\Sigma =\Sigma_1\bigotimes\Sigma_2$ between a UN $\Sigma_1$ across the 4 endpoints and a CS $\Sigma_2$ across the 3 periods on the same endpoint.  The estimated treatment differences and $\Sigma$ were then fed into the multivariate $t$-based power function to obtain the  power.    In the bootstrap approach, 1000 sets were generated and the hypothesis testing in each simulated $\tilde{\y}$  was the same as in the BBS and BS2 procedures.  The success rate out of the 1000  trials was  the bootstrapped power.

Though the  pilot study is a 3-period crossover study while the future study is 2-period, the model-based power and BEP values calculated above were legitimate since the treatment differences were adjusted for the period effects through the lme models fitted to the pilot data (model-based) or to the future data (bootstrap-based); and it was not necessary to produce the ``exact'' 2-period  crossover data in the FTS approach.  In other words, the design inconsistency between the pilot and future trials was not a concern from the perspective of the treatment comparison as along as the period effects were properly taken care of.

The results are presented in Table \ref{tab:case1} and summarized as follows.  First,  the BEP values via the BBS and BS2 procedures were similar.  Second,  the model-based BEP and power values were notably sensitive to the underlying parametric assumptions. The  assumptions on the  dependency structure  $\Sigma$ among the endpoints influenced the estimates on the treatment differences in the pilot data, leading different  power and BEP values under each assumption. Specifically, for BEP, different priors on $\Sigma$ led to a $\sim\!25\%$ difference; in the case of power, different  structural assumptions on $\Sigma$ led to a $25\%\!\sim\!35\%$ difference. Third,  the parametric and bootstrapped POS estimates were different in most cases.  Fourth,   T1 had consistently higher BEP than T2 though the absolute values varied by the assessment approach.  The relative POS of T1 and T2 measured by the classical power were slightly inconsistent  depending on the assessment approach. 
All taken together, we would recommend T1 for the future trial and cite the BEP obtained via the BBS and BS2 procedures given their robustness (model-free incorporation of the uncertainty around the parameters in the power calculation).
\vspace{-18pt}
\begin{center}\label{fig:sim1CDF}
\begin{table}[htb]
\centering
\begin{tabular}{p{2.0cm} p{1.3cm}p{1.3cm}p{0.8cm}p{0.8cm}p{0.2cm}p{1.3cm}p{1.3cm}p{1.0cm}}
\hline
& \multicolumn{4}{c}{BEP (\%)} && \multicolumn{3}{c}{Power (\%)}\\
\cline{2-5}\cline{7-9}	
treatment 	& model-  &  model- & BBS & BS2 &&  model- &  model-   & BS\\
                     &based$_1^{\dag}$ &  based$_2^{\dag}$ && && based$_1^{\ddag}$  & based$_2^{\ddag}$ &\\
\hline
T1 				& 51.0	&74.3  & 86.3 & 87.6 & & 99.98 & 65.8 	& 100.0\\
T2 				&	42.6	&68.0 & 79.4 & 81.6 & & 99.39 & 78.6 	& 99.3 \\
\hline
\end{tabular}
\footnotesize
\begin{tabular}{l}
 $^{\dag}$  Prior $\!f\!(\Sigma)\!\propto\!|\Sigma|^{-\frac{p+1}{2}}$ in parametric$_1$; $\!\propto\!|\Sigma|^{-(p+1)}$ in parametric$_2$   ($p\!=\!12$) \\
$^{\ddag}$ A fully parameterized $\Sigma_{p\times p}$ ($p\!=\!12$) was assumed on the pilot data in  parametric$_1$,  and\\
  $\Sigma\!=\!\Sigma_{\mbox{\tiny{UN}}}\!\otimes\!\Sigma_{\mbox{\tiny{CS}}}$ was  assumed in  parametric$_2$  ($\Sigma_{\mbox{\tiny{UN}}}$ was  fully parameterized across 4 endpoints,\\  and $\Sigma_{\mbox{\tiny{CS}}}$  followed  a  CS structure across the 3 repeated measures per endpoint).\\
\hline
\end{tabular}
\caption{POS assessment of each test formulation in the future BE study }\label{tab:case1}
\end{table}
\end{center}
\vspace{-18pt}

\subsection{example 2: joint modeling of survival time and longitudinal data}
 The pilot study is a randomized trial in 40  AIDS patients  who had failed or were intolerant of zidovudine (AZT) therapy (simulated from a published data set in \citet{aids}). The  patients were allocated  to two  antiretroviral drug treatment groups  (A and B) in a 1:1 ratio. Both the time to death due to AIDS and the CD4 count at 4  times points (0, 2, 6 and 12 months) were collected in the pilot study.  Each patient was followed up for a minimum of 12 months and with  an average of 15.6 months. Based on the results from the pilot study, a larger study of size $\tilde{n}$ is under planning to compare the two treatments in a more confirmatory manner and to test the hazard ratio of death between the two against 1. The team is interested in the assessment of the POS of the future trial.

In the model-based assessment of the BEP, the Bayesian joint modelling of the survival time  (with right censoring) and the CD4 counts was applied. Specifically, the square root of the CD4 count $y_{ij}$ in subject $i$ at time $t_j$ was modeled with a lme model after the square-root transformation; and the Cox model was applied to analyze the survival time  with the conditional expected  mean of $y_{ij}$ as a covariate, plus the effect from the treatments (Eqn \ref{eqn:jm}).
\begin{align}
y_{ij}(t) & = \eta_{ij} + \epsilon_{ij}(t) = \mathbf{x}_{ij}(t) \bs{\beta} +\mathbf{z}_{ij}(t)\mathbf{b}_i +\epsilon_{ij},\notag\\
h_i(t)& =  h_0(t) \exp ( \gamma x_{i2} + \lambda\eta_{ij}(t)),\label{eqn:jm}\\
\log(h_0(t)) &= \gamma_0 + \textstyle\sum_{l=1}^L \gamma_lB_l(t, \mathbf{k}),\notag
\end{align}
$\mathbf{x}_{ij}(t)$ is the covariates for patient $i$ at time $t_j$ that includes an intercept, the time when $y_{ij}$ was measured ($x_{ij1}(t)$),  the dummy variables for treatment  $x_{i2}$,  whether there was previous AIDS diagnosis at the study entry  ($x_{i3}$), and whether the patient had AZT intolerance or AZT failure ($x_{i4}$); $\bs{\beta}$ contains the fixed effects associated with $\mathbf{x}_{ij}(t)$; the random effect $\mathbf{b}_i=(b_{i0},b_{i1})' \sim N(\mathbf{0}, \mathbf{G}_{2\times2})$ corresponds a random intercept and a random slope for $x_{ij1}(t)$; and error term $\epsilon_{ij}\sim N(0, \sigma^2)$. $\lambda$ in Eqn (\ref{eqn:jm}) quantifies the association between the CD4 count  up to time $t$ and the hazard for death at  $t$, and $\gamma$ is the log hazard ratio for death between the two treatments. The baseline hazard $h_0(t)$ was modelled with the penalized-splines approach, where  $B_l(t, \mathbf{k})$ is the $l\textsuperscript{th}$ basis function of the splines at knots $\mathbf{k}=(k_1,\ldots,k_L)$ (we set $L=8$) and $\gamma_l$'s are the spline coefficients.  The R package \texttt{JMbayes} was employed to fit the joint model in Eqn (\ref{eqn:jm}). After the convergence of the MCMC algorithm, 2000 posterior samples on the model parameters, including $\mathbf{b}_i$, were obtained (after 1000 burn-in and 40 thinning), from which the posterior probability of survival $p_i=\Pr(t_i\le T)$ by some time $T$  in subject $i$ was obtained for $i=1,\ldots,40$. We examined two types of $T$:  the last recorded time in the pilot study, which was 20.87 and 20.27 months in Treatments A and B, respectively; and  24 months, which was an extrapolation of the pilot data in study duration. The model-based  power for testing  $H_0:$ HR $=1$ is 
$\Phi\!\left(\!\sqrt{\tilde{n} (1-p)\gamma^2}/2-\Phi^{-1}(1-\alpha/2)\right),$
where $\tilde{n}(1-p)$ is the expected number of events in the future trial, where $p=n^{-1}\sum_{i=1}^{n}p_i$,  and $\alpha$ is the type-I error rate. Power was calculated at each of the 2000 posterior  samples of  ($p, \gamma)$, the average of which gave the model-based BEP. The model-based power given the posterior means of $\gamma$ and $p$ at two types of $T$ was also calculated.

In the bootstrap-based approaches,  $m=500$ sets of future data were generated according to the steps in Table \ref{tab:bbs1} and  Table \ref{tab:bbs3} ($t=1$).  In each simulated trial, the joint model in Eqn (\ref{eqn:jm}) was applied, and the 95\%  posterior interval on the HR was obtained. If  the lower bound of the interval was $>1$ or the upper bound was $<1$,  then the future trial was claimed a success; the success rate out of 500 was the BEP. The bootstrap-based power based on Table \ref{tab:bbs3} was also calculated.  Different from the model-based approach, extrapolation to a longer study duration ($T=24$) was not possible without further parametric assumptions. In other words, the BEP and power evaluated via the bootstrap-based approaches  implicitly assumed the study duration in the future trial was the same as that in the  pilot study. 

The results of POS at different $\tilde{n}$ are provided in Table \ref{tab:case2}.  The model-based BEP and the BEP via the BBS and BS2 procedures were similar in this example, so were between the model-based  power and the bootstrap power.  In summary, the POS for the future trial reached $80\sim85\%$ at $\tilde{n}=400$, which might be an acceptable level of POS. As expected, a longer study duration (24 months) led to an increase in the POS (by $5\%\sim10\%$) compared to the same sized future trial with the same study duration as the pilot study. By contrast, all the POS values assessed by the power were close to 100\% at all $\tilde{n}$ and for both study durations, thus not a differentiable nor a valuable metric in determining $\tilde{n}$ .
\vspace{-18pt}
\begin{center}
\begin{table}[htb]\centering
\begin{tabular}{c ccc c cc c cc}
\hline
$\tilde{n}$ & \multicolumn{6}{c}{$T$ was the same as pilot study}  & &\multicolumn{2}{c}{$T=24$ months } \\
\cline{2-7} \cline{9-10}	
allocation & \multicolumn{3}{c}{BEP(\%)} && \multicolumn{2}{c}{Power (\%)} && BEP(\%) & Power$^{\dag}$(\%)\\
  ratio 1:1  & model-based  &  BBS & BS2 &&  model-based$^{\dag}$   & BS &&  \multicolumn{2}{c}{ model-based}\\
\cline{1-4}	\cline{6-7}	\cline{9-10}	
200	 &73.4 & 75.0 & 73.6  && 99.6    & 90.4 && 83.0  & $>$99.9 \\
400 &81.8  & 83.0 & 84.5 && $>$99.9 & 99.2 && 88.3& 100.0 \\
600 &85.5 & 87.0 & 86.5  && 100.0 & 99.8 &&  90.5 & 100.0 \\
\hline
\end{tabular}
\footnotesize
\begin{tabular}{l}
 $^{\dag}$  The model-based power was also calculated with the sample survival probability (57.5\%)  and  the  \\
   \hspace{8pt}posterior  mean of log-HR from the Bayesian joint model in the pilot data; the results were similar.\\
\hline
\end{tabular}
\caption{Empirical POS assessment for the future trial to compare two antiretroviral drugs in hazard of death in HIV patients}\label{tab:case2}
\end{table}
\end{center}
\vspace{-18pt}

\section{Discussion}\label{sec:discussion}
We developed the BBS procedure based on the Bayesian bootstrap to calculate the BEP  given individual-level pilot data. We also presented a non-Bayesian counterpart  to the BBS procedure, named the  double bootstrap (BS2), that achieves the same goal as the BBS  for the BEP assessment.  Neither procedures make assumptions about the distribution of the pilot data and can  handle multidimensional data sets without imposing a dependence structure among the variables. The implementation of both procedures are straightforward: only a few lines of codes are needed  and the whole  procedures can be easily parallelized for fast computation.  By contrast, the parametric approaches specify a likelihood function on the pilot data set and priors for the model parameters, followed by posterior sampling of the parameters involved in the analytical power calculation.  The derivation of the posterior distributions and the posterior sampling  can be complicated and computational costly, not to mention the possibility of  model mis-specification on the pilot data.

In the case of existence of multiple sets of pilot data, the data sets can be easily combined to simulate the future trial. The weight associated with each pilot study, in terms of their contribution to the future data, by default would be the size of the pilot studies: the larger the sample size of a pilot study, the more likely the subjects from that study will be  bootstrapped. If other weights are desired and specified (e.g., the design agreement between the pilot and future trials, the quality of the pilot data), they can be  conveniently incorporated using a 2-stage sampling procedure: first sampling the studies with probabilities  proportional to the corresponding weights and followed by the regular BBS and BS2 procedure within each study.

The bootstrap-based approaches  require the pilot information to be available in the form of individual-level data. When there exist only historical  summary/aggregate statistics, the bootstrap-based approaches are not directly applicable. Though future data can be simulated from the aggregate statistics, it would require additional distributional assumptions, defeating the purposes of developing the nonparametric bootstrap methods in the first place. In practice, the pilot study might not be perfectly matched up  to simulate the future trial such as the design discrepancy between the two. Some design differences may not have direct impacts on the planned  analysis on the future trial or can be adjusted for in the BEP calculation, and thus will not be a concern (e.g., the first case study). If a discrepancy that relates to the effect size in the power calculation is not easy to adjust for without making further parametric assumptions (such as the second example where the study duration could be different), the model-based approaches might be the only choice. However, if unreasonable extrapolation has to made, the parametric approaches would not be appropriate either. 

In summary, the BBS and BS2 approaches provide an alternative to the parametric approaches to assess the BEP. The bootstrap-based procedures will appeal to non-Bayesian practitioners  given their analytical and computational simplicity and easiness in implementation. We provide the sample R functions on the BBS and BS2 procedures  in the online supplementary materials to facilitate their practical implementation. 

\section*{\normalsize{Supplementary materials}}
\noindent The supplementary materials can be found at \url{http://www3.nd.edu/~fliu2/BBS-supp.pdf}

\bibliographystyle{apalike}

\appendix
\section*{\normalsize Appendix A: Conditional posterior distributions of $\Sigma$  and $\bs{\beta}$ in case study 1}
\noindent This appendix presents the derivation of the posterior distribution of $\Sigma$ and $\bs{\beta}$ in linear mixed models. Denote the number of subjects by $n$ and the number of measurements per subject by $p$. The likelihood of $\Sigma$ and $\bs{\beta}$ is
$L(\Sigma,\bs{\beta}|\y)\!\propto\!\textstyle\prod_{i=1}^n\!\left\{\!|\Sigma|^{-1/2}\mbox{exp} \left(-\frac{1}{2}(\y_i-\mathbf{x}_i\bs{\beta})^T\Sigma^{-1}(\y_i-\mathbf{x}_i\bs{\beta})\right)\!\right\}$.
Let the prior on $\bs{\beta}$ and $\Sigma$ be $p(\bs{\beta},\Sigma)\!=\!|\Sigma|^{-(\nu_0+p+1)/2}$.  The full conditional posterior distribution of $\Sigma$ given $(\bs{\beta},\y)$ is
\begin{align*}
p(\bs{\Sigma}|\bs{\beta},\y)
&\propto |\Sigma|^{-(\nu_0+p+n+1)/2} \mbox{exp}\left(\textstyle-\frac{1}{2}\sum_{i=1}^n(\y_i-\mathbf{x}_i\bs{\beta})^t\Sigma^{-1}(\y_i-\mathbf{x}_i\bs{\beta})\right)\\
&\propto |\Sigma|^{-(\nu_0+p+n+1)/2}\mbox{exp}\left(\textstyle-\frac{1}{2}\sum_{i=1}^n\mbox{tr}(\mathbf{e}_i\mathbf{e}_i^T\Sigma^{-1})\right)\\
&\propto |\Sigma|^{-(\nu_0+p+n+1)/2}\mbox{exp}\left(\textstyle-\frac{1}{2}\mbox{tr} \left(\sum_{i=1}^n\mathbf{e}_i\mathbf{e}_i^T\Sigma^{-1}\right)\right), \mbox{ where } \mathbf{e}_i=\y_i-\mathbf{x}_i\bs{\beta}.\\
\mbox{Thus } p(\bs{\Sigma}|\bs{\beta},\y)&=\frac {|\Sigma|^{-(\nu_0+p+n+1)/2}|\sum_{i=1}^n\mathbf{e}_i\mathbf{e}_i^t|^{n/2} \mbox{exp}\left(-\mbox{tr}\left(\sum_{i=1}^n\mathbf{e}_i\mathbf{e}_i^t\Sigma^{-1}\right)/2\right)}{2^{(mn/2)}\Gamma_n(m/2)},
\end{align*}
which is an inverse Wishart distribution $W^{-1}(\sum_{i=1}^n\mathbf{e}_i\mathbf{e}_i^t, n)$ with scale matrix $\sum_{i=1}^n\mathbf{e}_i\mathbf{e}_i^t$ and the degrees of freedom $n$. The full conditional distribution of $\bs{\beta}$ given $(\Sigma,\y)$ is
\begin{align*}
p(\bs{\beta} |\bs{\Sigma},\y)
&\propto |\Sigma|^{-(\nu_0+p+n+1)/2}\mbox{exp}\left(\textstyle-\frac{1}{2}\sum_{i=1}^n(\y_i-\mathbf{x}_i\bs{\beta})^T\Sigma^{-1}(\y_i-\mathbf{x}_i\bs{\beta})\right)\\
&\propto |\Sigma|^{-(\nu_0+p+n+1)/2}\mbox{exp}\left(\textstyle-\frac{1}{2}\sum_{i=1}^n \left(\Sigma^{-1/2}(\y_i-\mathbf{x}_i\bs{\beta})\right)^T\left(\Sigma^{-1/2}(\y_i-\mathbf{x}_i\bs{\beta})\right)\right)\\
&\propto |\Sigma|^{-(\nu_0+p+n+1)/2}\mbox{exp}\left(\textstyle-\frac{1}{2}
\left(\bs{\beta}^T (\sum_{i=1}^n\x_i^T\Sigma^{-1}\x_i)\bs{\beta} \right.\right.\\
&\left.\left.\qquad\qquad\qquad\textstyle-\sum_{i=1}^n \y_i^T\Sigma^{-1}\x_i\bs{\beta}-\bs{\beta}^T\sum_{i=1}^n \x_i^T\Sigma^{-1}\y_i+\sum_{i=1}^n\y_i^T\Sigma^{-1}\y_i\right)\right)\\
\propto |\Sigma|^{-(\nu_0+p+n+1)/2}&\mbox{exp}\left(\textstyle-\frac{1}{2}
\left(\bs{\beta}^T (\x^T\Omega^{-1}\x)\bs{\beta} - \y^T\Omega^{-1}\x\bs{\beta}-\bs{\beta}^T \x^T\Omega^{-1}\y+\y^T\Omega^{-1}\y\right)\right)\\
\propto |\Sigma|^{-(\nu_0+p+n+1)/2}
\end{align*}\vspace{-28pt}
$$\mbox{exp}\left(\textstyle-\frac{1}{2}
\left(\bs{\beta} - (\x^T\Omega^{-1}\x)^{-1}(\x^T\Omega^{-1}\y)\right)^T\left((\x^T\Omega^{-1}\x)^{-1}\right)^{-1}
\left(\bs{\beta} - (\x^T\Omega^{-1}\x)^{-1}(\x^T\Omega^{-1}\y)\right)\right).$$
Thus $p(\bs{\beta} |\bs{\Sigma},\y)=  \mbox{N}\left((\x^T\Omega^{-1}\x)^{-1}(\x^T\Omega^{-1}\y), (\x^T\Omega^{-1}\x)^{-1}\right)$,
where $\Omega$ is a block diagonal matrix with $n$ blocks of $\Sigma$ on the diagonal, $\x=\{\x_i\}_{i=1,\ldots,n}$, and
$\y=\{\y_i\}_{i=1,\ldots,n}$

\end{document}